\documentclass{article}

\usepackage{microtype}
\usepackage{graphicx}
\usepackage{booktabs} 
\usepackage{amsmath}
\usepackage{amssymb}
\usepackage{amsfonts}
\usepackage{soul}
\usepackage{physics}
\usepackage{subcaption}
\usepackage{multirow}

\usepackage{hyperref}

\usepackage{algorithm}
\usepackage{algorithmic}
    
\usepackage{graphicx}
\usepackage{xcolor}
\usepackage{verbatim}
\usepackage{tikz}
\usetikzlibrary{spy}
\usepackage{textcomp}
\usepackage{adjustbox}
\usepackage{siunitx}
\usepackage{array}
\usepackage{commath}
\usepackage{url}
\usepackage{paralist}
\usepackage{booktabs} 
\usepackage{gensymb}
\usepackage{enumitem}
\usepackage{cancel}
\setlist{nolistsep}
\graphicspath{ {./Figs/} }

\usepackage{wrapfig}

\newcommand\blfootnote[1]{%
  \begingroup
  \renewcommand\thefootnote{}\footnote{#1}%
  \addtocounter{footnote}{-1}%
  \endgroup
}

\usepackage[accepted]{mlsys2024}

\mlsystitlerunning{Ultra-Long Sequence Distributed Transformer}

\begin{document}

\twocolumn[
\mlsystitle{Ultra-Long Sequence Distributed Transformer}




\begin{mlsysauthorlist}
\mlsysauthor{Xiao Wang}{ORNL}
\mlsysauthor{Isaac Lyngaas}{ORNL}
\mlsysauthor{Aristeidis Tsaris}{ORNL}
\mlsysauthor{Peng Chen}{AIST}
\mlsysauthor{Sajal Dash}{ORNL}
\mlsysauthor{Mayanka Chandra Shekar}{ORNL}
\mlsysauthor{Tao Luo}{ASTAR}
\mlsysauthor{ Hong-Jun Yoon}{ORNL}
\mlsysauthor{Mohamed Wahib}{RIKEN}
\mlsysauthor{John Gounley}{ORNL}
\end{mlsysauthorlist}

\mlsysaffiliation{ORNL}{Oak Ridge National Lab, US}
\mlsysaffiliation{RIKEN}{RIKEN Center for Computational Science, Japan}
\mlsysaffiliation{AIST}{The National Institute of Advanced Industrial Science and Technology, Japan}
\mlsysaffiliation{ASTAR}{Agency for Science, Technology and Research, Singapore}
\mlsyscorrespondingauthor{Xiao Wang}{wangx2@ornl.gov}

\mlsyskeywords{Transformer, Sequence Parallelism, Large Language Model}

\vskip 0.3in

\begin{abstract}
Transformer models trained on long sequences often achieve higher accuracy than short sequences. Unfortunately, conventional transformers struggle with long sequence training due to the overwhelming computation and memory requirements. Existing methods for long sequence training offer limited speedup and memory reduction, and may compromise accuracy. This paper presents a novel and efficient distributed training method, the Long Short-Sequence Transformer (LSS Transformer), for training transformer with long sequences.
It distributes a long sequence into segments among GPUs, with each GPU computing a partial self-attention for its segment. Then, it uses a fused communication and a novel double gradient averaging technique to avoid the need to aggregate partial self-attention and minimize communication overhead. We evaluated the performance between LSS Transformer and the state-of-the-art Nvidia sequence parallelism on a Wikipedia \textit{enwik8} dataset. Results show that our proposed method lead to 5.6x faster and 10.2x more memory-efficient implementation compared to state-of-the-art sequence parallelism on 144 Nvidia V100 GPUs. Moreover, our algorithm scales to an extreme sequence length of 50,112 at 3,456 GPUs, achieving  161\% super-linear parallel efficiency and a throughput of 32 petaflops.

\end{abstract}
]




\section{Introduction}
\label{sec:prob_overview} 

\blfootnote{1, Oak Ridge National Laboratory, US. 2, National Institute of Advanced Industrial Science and Technology, Japan. 3, Agency for Science, Technology and Research, Singapore. Corresponding email: wangx2@ornl.gov}The transformer is a powerful neural network architecture widely used in natural language and image processing~\cite{Vaswani17}. Its versatility is evidenced by its wide range of applications, including machine translation~\cite{transformer-wang19}, chatbots~\cite{caldarini2022literature}, speech recognition~\cite{dong2018speech}, image captioning~\cite{yu2019multimodal}, image segmentation~\cite{valanarasu2021medical,strudel2021segmenter}, and classification~\cite{chen2021crossvit}.
The transformer achieves its impressive performance by recognizing that different input sequence tokens have varying levels of importance to the final output prediction. The transformer captures the relationship between each pair of input tokens using a process called {\em ``self-attention''}. This allows the transformer to generate highly accurate outputs by focusing on the most relevant tokens in an input sequence while also paying attention to the overall context. This approach has proven to be highly effective and makes transformer a leading technology in artificial intelligence.

\begin{table*}[t]
\caption{A summary for different long sequence training methods. \emph{Serial} methods may need to be used on top of parallelism schemes. \emph{$l_x$}= sequence length; \emph{N}= number of GPU workers; \emph{H}= number of hierarchical levels; \emph{Z}= number of non-zeros.}
\label{tab:related_work}
\resizebox{\linewidth}{!}{
\centering
\begin{tabular}{|c|l|l|l|l|l|l|l|l|l|} 
\hline
\textbf{Approach}                                                                       & \textbf{Method}
& \multicolumn{1}{c|}{\begin{tabular}[c]{@{}c@{}}\textbf{Accuracy}\\\textbf{~Loss}\end{tabular}}                                                                        
& \multicolumn{1}{c|}{\begin{tabular}[c]{@{}c@{}}\textbf{Serial/}\\\textbf{~Distributed}\end{tabular}}
& \multicolumn{1}{c|}{\begin{tabular}[c]{@{}c@{}}\textbf{Memory}\\\textbf{~(per worker)}\end{tabular}} 
& \multicolumn{1}{c|}{\begin{tabular}[c]{@{}c@{}}\textbf{Compute}\\\textbf{~(per worker)}\end{tabular}} 
& \multicolumn{1}{c|}{\begin{tabular}[c]{@{}c@{}}\textbf{Memory}\\\textbf{Total}\end{tabular}} 
& \multicolumn{1}{c|}{\begin{tabular}[c]{@{}c@{}}\textbf{Compute}\\\textbf{Total}\end{tabular}} 
& \multicolumn{1}{c|}{\begin{tabular}[c]{@{}c@{}}\textbf{Comm.}\\\textbf{freq. per layer}\end{tabular}}                                                                        
& \multicolumn{1}{c|}{\begin{tabular}[c]{@{}c@{}}\textbf{Practical}\\\textbf{~bottleneck}\end{tabular}} 
\\
\hline
\begin{tabular}[c]{@{}c@{}}\textbf{Hierarchical }\\\textbf{Training}\end{tabular}
&\begin{tabular}[c]{@{}l@{}}\cite{Chen22,chen2021crossvit}\\ \cite{si21, yu2023megabyte}\end{tabular} &
    No       & 
    Serial  &
    O($l_x^2H$)                             & 
    O($l_x^3H$)                                    & 
    N/A                             & 
    N/A                                    & 
    N/A &
    \begin{tabular}[c]{@{}l@{}}Train multiple \\models \end{tabular}\\ 
\hline
\begin{tabular}[c]{@{}c@{}}\textbf{Attention }\\\textbf{Approximation}\end{tabular}
    &\begin{tabular}[c]{@{}l@{}} \cite{reformer20,Roy21} \\ \cite{sparse-transformer19,Beltagy2020Longformer}\end{tabular} &
    Yes       & 
    Serial  &
    O($Z$)                             & 
    O($Z^3$)                                    & 
    N/A                             & 
    N/A                                    & 
    N/A &
    Low sparsity\\ 
\hline
\multirow{3}{*}{\begin{tabular}[c]{@{}c@{}}\textbf{Sequence }\\\textbf{Parallel}\end{tabular}} 
    & \begin{tabular}[c]{@{}l@{}}Straightforward Seq. Parallel\\ \cite{Li21,li2023lightseq}  \end{tabular} & 
    \multirow{4}{*}{No}       & 
    \begin{tabular}[c]{@{}l@{}}Distributed \\ (partial aggr.)  \end{tabular}  &
    O(${l_x^2} / {N}$)                             & 
    O(${l_x^3} / {N}$)                                    & 
    O($l_x^2$)                             & 
    O($l_x^3$)                                    & 
    $\approx N^2$ &
    \begin{tabular}[c]{@{}l@{}}Frequent \\ communication  \end{tabular}
    \\ 
\cline{2-2}\cline{4-10}
    & \begin{tabular}[c]{@{}l@{}}Nvidia Seq. Parallel\\ \cite{Korthikanti22}  \end{tabular}  & 
           & 
    \begin{tabular}[c]{@{}l@{}}Distributed \\ (serial attention)  \end{tabular}   &
    O($l_x^2$)                             & 
    O($l_x^3$)                                    & 
    O($l_x^2$)                             & 
    O($l_x^3$)                                    & 
    $8$ &
    \begin{tabular}[c]{@{}l@{}}Attention not \\ distributed \end{tabular} \\
\cline{2-2}\cline{4-10}
    & \textbf{LSS Transformer (ours)}& 
    & 
    \begin{tabular}[c]{@{}l@{}}Distributed \\ (fully) \end{tabular}    &
    O(${l_x^2} / {N}$)                             & 
    O(${l_x^3} / {N}$)                                    & 
    O($l_x^2$)                             & 
    O($l_x^3$)                                    & 
    $2$ &
    - \\
\hline
\end{tabular}
}
\end{table*}

With long sequence training, transformer attends to many more input tokens than a transformer trained with short sequences. Therefore, long sequence training often captures more contextual information and leads to markedly higher prediction accuracy for many tasks with long-range dependencies, such as DNA sequence analysis~\cite{bigbird20}, long document summary~\cite{Beltagy2020Longformer} and image segmentation~\cite{strudel2021segmenter}. Unfortunately, transformer's memory footprint increases quadratically and computations increase cubically with longer sequence lengths~\cite{Beltagy2020Longformer,FlashAttention22}. Therefore, the transformer's sequence length is typically truncated to no more than a couple thousand tokens due to runtime and memory constraints, despite longer sequences leading to higher accuracy. 

To address this issue, there are three research approaches: hierarchical training, attention approximation, and distributed sequence parallelism. Hierarchical training involves training multiple transformers at different levels of abstraction~\cite{si21,Chen22,chen2021crossvit, yu2023megabyte}. The transformer at the lowest abstraction level trains on the shortest sequence segments. Then, the transformer at the next higher level uses the previous level outputs as additional input to train on longer segments. The process then repeats until reaching the highest abstraction level.  However, training multiple transformers at different abstraction levels significantly increases training time and memory footprint. In addition, hyper-parameter tuning is required for the optimal model architecture at each abstraction level.

The approximation approach, in contrast, aims to reduce the computations and memory usage by approximating the self-attention operations through either sparse sampling~\cite{sparse-transformer19,reformer20,Roy21,Beltagy2020Longformer,bigbird20}, low-rank approximation~\cite{Choromanski20,Katharopoulos20}, infrequent self-attention updates~\cite{ying2021lazyformer,rabe2022selfattention}, or their combinations~\cite{chen2021scatterbrain}.  These approximation methods can significantly reduce memory footprint and computations, and some can even reduce the long sequence problem's quadratic complexity to linear complexity. However, approximation is a lossy information compression technique that discards partial information for the self-attention. Thereby, excessive approximation may lower accuracy especially for sequences with long-range dependency. Previous experiments demonstrate significant accuracy degradation when approximation compression ratio exceeds 70\%~\cite{Shi21}.

The distributed sequence parallelism approach aims to address the long sequence problem by distributing long sequences into contiguous sequence segments among GPUs~\cite{Li21,li2023lightseq, Korthikanti22, jacobs2023deepspeed}. The largest hurdle in distributed sequence parallelism lies in handling the self-attention, which is both the most compute-intensive but also the most communicate-intensive step. Since each GPU's sequence segment has a correlation with every other GPU's segment, a straightforward distributed self-attention method, such as~\cite{Li21}, Deep-Speed Ulysses~\cite{jacobs2023deepspeed} and the most recent LightSeq~\cite{li2023lightseq}, requires each GPU to communicate with every other GPU multiple times to calculate a partial self-attention score for its assigned segment, before aggregating them into the final self-attention output. Consequently, communication frequency tends to increase significantly in a quadratic growth rate with more sequence parallel GPUs, substantially impeding scalability. 

Conversely, an alternative sequence parallel method from Nvidia's Megatron-LM framework completely avoids self-attention dependency by performing sequential updates on the core self-attention and feed-forward operations along the sequence dimension, while parallelizing less compute-intensive but independent tasks, such as layer normalization and dropouts~\cite{Korthikanti22}. Although this method requires only 8 communications per attention layer, it misses the opportunity to parallelize the compute intensive self-attention. Consequently, it leads to a modest 29\% speedup compared to a sequential baseline with full backward recomputation~\cite{Korthikanti22}.

A summary of each approach's core idea and limitations is provided in Table~\ref{tab:related_work}. To address their limitations, this paper introduces the ``Distributed Long Short-Sequence Transformer" (LSS Transformer).
In contrast to existing methods, the LSS Transformer (1) utilizes a single transformer for training; (2) attains remarkable speedup and memory reduction; (3) has no approximation, thereby no accuracy loss; and (4) maintains a minimal communication overhead, requiring only 2 communications per attention layer.
It distributes a long sequence into short segments among GPUs, with each GPU computing a partial self-attention score for its segment in the context of the entire sequence. Then it uses a fused communication and a double gradient averaging technique to avoid the need to aggregate partial self-attention score, minimize communication overhead while preserving data dependency. 

This paper contributes in the following ways: (1) Introducing a general and innovative end-to-end sequence framework for long sequence training. Operating at the attention layer level, it remains agnostic to model sizes and variations (encoder-only, decoder-only, etc.), making it universally applicable without modifications. (2) Presenting a low communication overhead algorithm that uses fused communication and gradient averaging to lower communication frequency. (3) Presenting an integrative sequence and data parallelism algorithm that minimizes inter-GPU communications within local communicative groups. Our assessment on the Wikipedia \textit{enwik8} dataset demonstrates the superiority of the LSS Transformer over state-of-the-art Nvidia sequence parallelism~\cite{Korthikanti22}, achieving 6x faster training and 10x enhanced memory efficiency on 144 Nvidia V100 GPUs. Moreover, our algorithm scales remarkably to an extensive sequence length of 50,112 using 3,456 GPUs, delivering 161\% super-linear parallel efficiency and a computation throughput of 32 petaflops.

\begin{table*}[t]
\caption{Three distinct challenges for training transformer and  their orthogonal levels of parallelism. \emph{$N_d$} = data size; \emph{$N_m$} = model size; \emph{$l_x$} = sequence length; \emph{$B$} = batch size;}
\vskip 0.05in
\begin{center}
\begin{small}
\begin{tabular}{l|ccr}
\toprule
Challenges & Large Dataset & Large Model Size & Long Sequences \\
\midrule
Computational complexity & $O(N_d)$   & $O(N_m)$ & $O(l_x^3)$ \\
Memory complexity & $O(B)$   & $O(N_m)$ & $O(l_x^2)$ \\
Parallelism & Data Parallel   & Model Parallel & Sequence Parallel \\
Source of parallelism & Distributed input data batches   & Distributed model parameters & Distributed sequence segments \\
Distributed memory & No   & Yes & Yes \\
\bottomrule
\end{tabular}
\end{small}
\end{center}
\vskip -0.1in
\label{table:challenges}
\end{table*}

\section{Background and  Motivation}
\subsection{Orthogonal Levels of Parallelism}
\label{sec:orthogonal-parallelism}
Training a transformer model has three distinct computation challenges: (1) large training dataset, (2) large model size and (3) long sequences. Table~\ref{table:challenges} offers a concise overview of these challenges and their characteristics. As discussed earlier, increasing sequence length leads to a cubic growth rate for computations and a quadratic growth rate for memory use. In contrast, scaling model parameters or data batches presents linear computational and memory complexities. 

Each challenge relies on a unique parallelism. These parallelism levels are mostly orthogonal, address different challenges and cannot replace each other. Data parallelism accelerates training on large datasets by distributing training batches. However, it does not distribute memory and each GPU has a copy of the entire model. Model parallelism achieves speedup and distributed memory for large model problem by distributing model parameters and their gradients. Sequence parallelism, in contrast, accelerates computations and distributes memory for long sequences by distributing sequence segments. 
This paper parallelizes in the sequence dimension to scale the sequence length by distributing memory and compute cost, while also carefully assuring that the sequence parallelism does not interfere with or hinders other forms of parallelization.

\begin{figure*}[t]
\centering
\includegraphics[width=\linewidth]{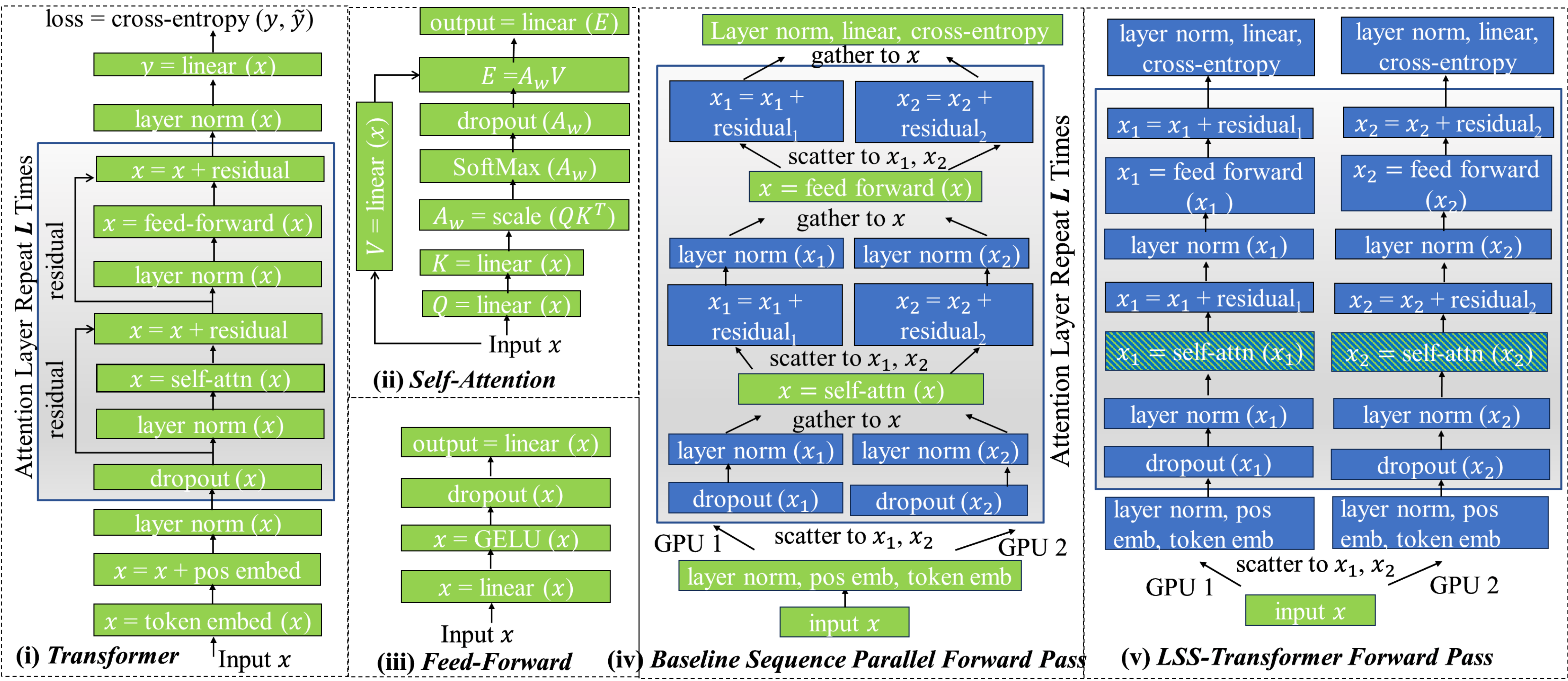}
\caption{(i) A generic transformer with $L$ attention layers outlined by a gray box. (ii) and (iii) the self-attention and the feed-forward units, respectively. (iv) Forward pass for baseline sequence parallelism with 2 sequence distributed GPUs. Blue indicate distributed steps and green for sequential steps. (v) The 2 GPU example for the LSS Transformer's forward pass.}
\label{fig:transformer-arch}
\end{figure*}

\subsection{Transformer Architecture}
\label{sec:transformer-arch}
Figs.~\ref{fig:transformer-arch}(i)-(iii) illustrate a standard transformer architecture for both the decoder only (i.e., GPT) and the encoder only (BERT) models. A token embedding in Fig.~\ref{fig:transformer-arch}(i) converts a tokenized image or text input sequence $x$ into an input vector. The input vector is then enhanced with embedded positional information. 

Next, the input vector $x$ of size $l_x \times E_m$, with $l_x$ representing the sequence length and $E_m$ being the embedding size, undergoes layer normalization and random dropouts before being processed by the self-attention. The detailed steps for self-attention are illustrated in Fig.~\ref{fig:transformer-arch}(ii) with the goal of producing a contextual embedding for each input token in relation to the whole sequence.
In the self-attention unit, the input vector $x$ is linearly transformed into query ($Q$), key ($K$), and value ($V$) vectors, with all three vectors having the same size as input $x$. Then self-attention computes the following equation:
\begin{equation}
E=\text{softmax} \left( QK^T / \sqrt{d_k} \right) V =A_w V \ .
\label{eqn:self-attention}
\end{equation}
The self-attention score, denoted as $A_w$, has size $l_x \times l_x$ and quantifies the correlation between each pair of tokens. This correlation is calculated by taking the dot product between $Q$ and the transposed $K$. To stabilize gradients during training, the self-attention score $A_w$ is further scaled by a constant $\sqrt{d_k}$, and is then normalized by a dropout and a SoftMax activation.
The final self-attention output $E$ is obtained by weighted averaging between value vector $V$ and $A_w$. The output embedding $E$ is then linearly transformed before exiting the self-attention unit.

The contextual embedding output from self-attention is further processed by a feed-forward unit, with detailed steps illustrated in Fig.~\ref{fig:transformer-arch}(iii), residual connections, and layer normalization to enhance training stability and prediction accuracy. These components, along with dropouts, form a single attention layer, outlined by a gray box in Fig.~\ref{fig:transformer-arch}(i). Then the attention layers repeat itself by $L$ times, where $L$ is the number of attention layers. After undergoing final layer normalization and linear transformation, the output $y$ is compared to the target $\tilde{y}$ using cross-entropy loss, and all parameters are updated. 

It's crucial to note that self-attention dominates both computation and memory requirements for training. With a size of $l_x \times l_x$, the self-attention score, $A_w$, has quadratic expansion in its size with increased sequence length. This leads to transformer's cubic computational and quadratic memory complexities with respect to sequence length.

\subsection{State-of-the-Art Sequence Parallelism}
\label{sec:baseline-seq-par}

To parallelize self-attention, the straightforward distributed self-attention method, such as~\cite{Li21,li2023lightseq}, partitions both the input vector $x$ and its linearly transformed vectors $Q$, $K$ and $V$ into distributed segments among GPUs. For example in a scenario with 3 GPUs, GPU 1 receives the first segments of these vectors, namely $x_1$, $Q_1$, $K_1$, and $V_1$, and GPUs 2 and 3 receive the second and the third segments. To compute Eqn.~(\ref{eqn:self-attention}), each GPU must receive every other GPU's segment to compute partial self attention scores. For example, GPU 1 needs to receive $K_2$ and $K_3$ from GPUs 2 and 3 before GPU 1 can compute partial self-attention scores $Q_1K_{2}^T$ and $Q_1K_{3}^T$. Then, the partial self-attention scores are aggregated across GPUs into the complete self-attention score, which is then used in dot products with the distributed value vector $V$. Since each GPU must communicate with every other GPU multiple times, the communication frequency for the straightforward distributed self-attention method increases quadratically with more GPUs, significantly limiting its scalability.

To address this limitation, Nvidia's method~\cite{Korthikanti22}, referred to as ``baseline sequence parallelism" in the rest of the paper, computes self-attention and feed-forward sequentially to avoid the quadratically increased cross-GPU communications~\cite{Korthikanti22}. However, it independently parallelizes layer normalization and dropouts in the sequence dimension, as they lack such inter-token dependencies~\cite{Korthikanti22}. Note that the term ``sequential computations" pertains to single GPU computations, while ``parallel computations" refer to those performed across GPUs in the sequence dimension. Although self-attention and feed-forward are computed within a single GPU, their computations are still vectorized through parallel threads within the GPU.

Fig.~\ref{fig:transformer-arch}(iv) summarizes the baseline sequence parallelism using an example of 2 GPUs. During a forward pass, positional and token embeddings are computed sequentially and the output is scattered into contiguous segments among the GPUs along the sequence dimension. Then in the attention layers,
the feed-forward and the self-attention are sequentially updated by a single GPU, and are represented by green rectangles in the figure. All other steps are independently updated with sequence parallelism, represented by blue rectangles. Gather and scatter communications are used before and after self-attention and feed-forward to ensure sequential updates for them and independent parallel updates for all other steps. Finally, the results from GPUs are gathered for a final layer normalization, linear transform, and cross-entropy loss evaluation. In the backward pass, all steps are the same as the forward pass, except the gather communications in the forward pass is replaced by reduce-scatter for gradient synchronization among GPUs and scatter communications in the forward pass are replaced with gather operations in the backward pass.

Despite that the baseline sequence parallelism avoid the frequent GPU communications in self-attention as in the straightforward distributed method, it has two main limitations. First, the most compute-intensive self-attention and feed-forward steps are sequential. Secondly, the communication overhead is still significant with 8 global communications per attention layer (4 in forward pass and 4 in backward pass). As a result, the baseline sequence parallelism achieves only 3\% speedup on a 22-billion-parameter model compared to a baseline without backward pass recomputation, and up to 29\% speedup compared to a baseline with backward recomputation~\cite{Korthikanti22}.

\section{The Distributed Long Short-Sequence Transformer}
\label{sec:LSS Transformer}
To achieve excellent scalability, the LSS Transformer must distribute self-attention, but also overcome the quadratically increased communication frequency. The LSS Transformer performs sequence parallelism based on these principles:

\noindent\ul{\textbf{Principle 1: Independent Computations with Distributed Memory.}}
Except for self-attention, all other computations like layer normalization, residual connection, dropout, feed-forward, linear transform, positional, and token embedding can be independently computed and distributed among GPUs without dependencies in the sequence dimension. Memory storage for these steps is also distributed in the sequence dimension in the same manner.

\begin{figure*}
\centering
\includegraphics[width=\linewidth]{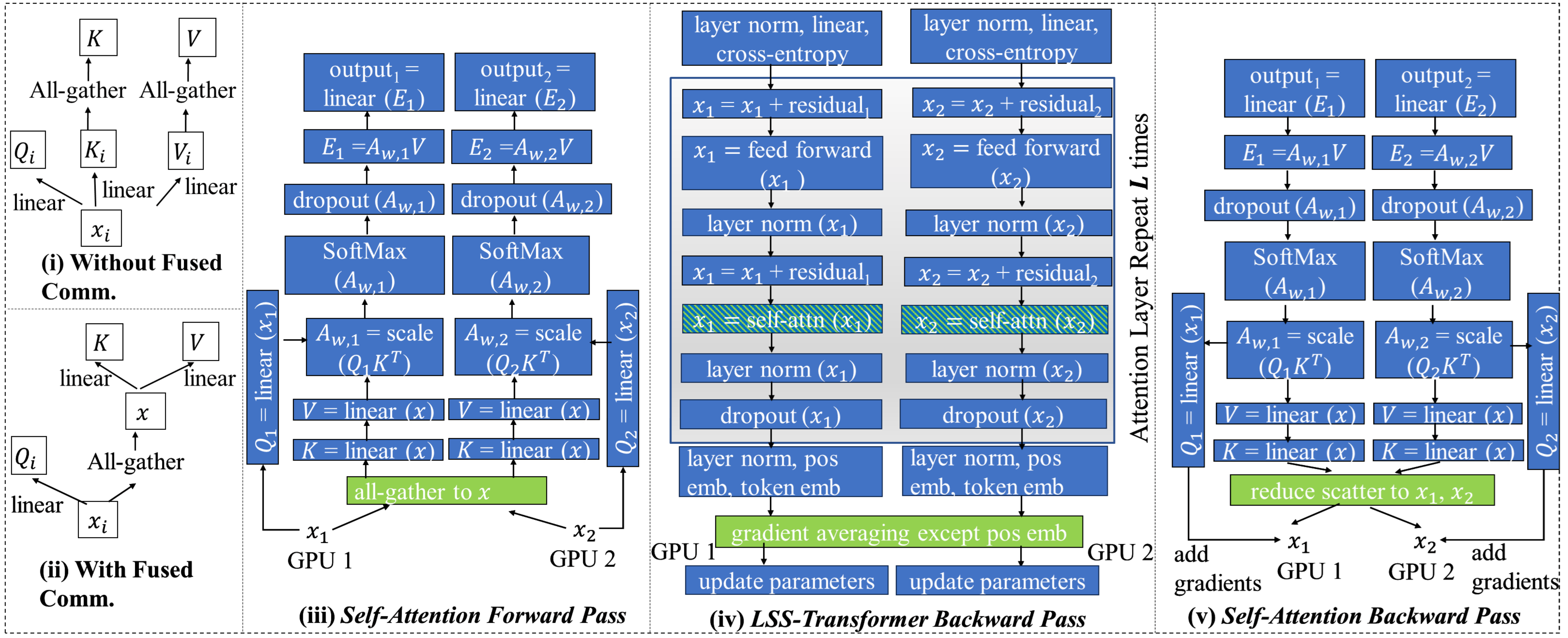}
 \caption{(i) and (ii) show the difference without and with fused communications. (iii) shows distributed self-attention's forward pass with fused communications. Note that the distributed self-attention outputs are not concatenated. (iv) LSS Transformer's Backward pass. Model parameters, except the positional embedding, are synchronized through gradient averaging. (v) The distributed self-attention's backward pass with reduce-scatter.}
\label{fig:LSS-Transformer} 
\end{figure*}

The feed-forward operation can be independently computed by row-distributing its linear transform multiplication along the sequence length dimension, enabling independent GPU computations. Additionally, GeLu activation and dropout operate independently on each element of their input. For instance, for a sequence input $x$ of dimension $l_x \times E_m$, where $l_x$ is sequence length and $E_m$ is embedding size, the first linear transform step in the feed-forward unit, $linear(x)$ in Fig.~\ref{fig:transformer-arch}(iii), multiplies $x$ with linear transform model parameters of size $E_m \times D_{inner}$, where $D_{inner}$ is the dimension of the feed-forward hidden layer. This matrix multiplication can be row-distributed among GPUs without communication when input $x$ is distributed into sequence segments $x_i$ of size $\frac{l_x}{N} \times E_m$, where $N$ is the number of sequence parallel GPUs. After independent element-wise operations on GeLu and dropout, the sequence distributed output is used as distributed input for the second linear transform multiplication in the feed forward unit, again row-distributed without communication.

Fig.~\ref{fig:transformer-arch}(v) depicts the LSS Transformer's forward pass, showcasing a demonstration with 2 sequence parallel GPUs. Note that while the figure exemplifies a specific scenario, the LSS Transformer's operate at the attention layer level, and is universally adaptable to various model sizes and types without modification.  In this figure, input sequence $x$ is scattered to $x_1$ and $x_2$ in the sequence dimension and each GPU receives a segment. Principle 1 enables all subsequent operations to be sequence distributed among GPUs and computed independently. In addition, each operations' inputs, intermediate outputs and their associated gradients are also stored in distributed memory across sequence-parallel GPUs, enabling excellent memory footprint reduction. We use blue rectangles in the figure to represent these independent computation steps. The self-attention is marked by a shaded blue rectangle to indicate that self-attention is distributed yet requires inter-GPU communications.

\noindent\ul{\textbf{Principle 2: Sequence Distributed Positional Embedding.}} The positional embedding parameters are a lookup table and each row of the table represents a token's spatial position within the sequence. The number of rows of the lookup table corresponds to the sequence length. Since each sequence-distributed GPU receives contiguous sequence segments, the GPU performs lookup operations only on the corresponding contiguous rows of the lookup table. This allows for row distribution of the positional embeddings among the GPUs without dependency.

\noindent\ul{\textbf{Principle 3: Distributed Self-Attention with Fused Communications.}}
To parallelize self-attention, we  use the following math property to preserve data dependency while minimizing communication overhead. By distributing the query vector, $Q$, in the sequence dimension among GPUs, we compute the self-attention output, $E$, as the following concatenation in the sequence dimension, and in parallel:
\begin{equation}
E= \begin{bmatrix} \text{softmax} \left( {Q_1}K^T \mathbin{/} \sqrt{d_k} \right) V \\ 
\text{softmax} \left({Q_2}K^T \mathbin{/}\sqrt{d_k} \right) V \\  \text{softmax} \left({Q_3}K^T \mathbin{/}\sqrt{d_k} \right) V \\ ... \end{bmatrix} \ , 
\label{eqn:self-attention-concat}
\end{equation}
where $Q_i$ is an $\frac{l_x}{N} \times E_m$ distributed query vector segment received by the $i^{\it th}$ GPU, where $N$ is the number of GPUs.  $V$ and $K$ are $l_x \times E_m$ collected value and key vectors without distribution, having the same copy across all GPUs. $\text{softmax} ({Q_i}K^T / \sqrt{d_k})$ represents the $i^{\it th}$ GPU's partial self-attention score for its assigned segment, but in the context of the whole sequence.

In summary, the key idea is to distribute the query vector in the sequence dimension among GPUs, while keeping the value and key vectors collected. Then LSS Transformer computes an individual self-attention output, $\text{softmax} ({Q_i}K^T / \sqrt{d_k})V$, for each GPU. Eqn.~(\ref{eqn:self-attention-concat}) shows that concatenating GPUs' individual self-attention outputs is \textbf{\em numerically the same} as directly computing self-attention $E$ sequentially, since the concatenation is a simple row gathering for $E$. Therefore, our proposed distributed self-attention method is an exact method without approximation, thereby no accuracy loss.
Compared to the straightforward sequence parallelism with quadratically increased communication frequency, a significant advantage of Eqn.~(\ref{eqn:self-attention-concat}) is that it enables distributed computing while requiring only 6 communications per self-attention layer. The forward pass requires 2 communications from gathering value and key vectors and 1 from self-attention concatenation. The backward pass requires another 3 communications.

To further reduce communication overhead, we use a fused communications technique to reduce the communication frequency from 6 communications per layer to 4. Fig.~\ref{fig:LSS-Transformer}(i) demonstrates the operations in the forward pass without the fused communication. An example sequence segment, $x_i$ is linearly transformed into query, key and value segments. Then, two all-gather communications are independently operated on the key and value segments into the collected $K$ and $V$. Fig.~\ref{fig:LSS-Transformer}(ii) shows the fused communication operation in the forward pass, requiring only a single all-gather communication. $x_i$ is linearly transformed into query segment $Q_i$. Meanwhile, $x_i$ is gathered into a temporary collected sequence $x$, before $x$ is linearly transformed into the collected key and value vectors. The same technique is also applied to backward pass, reducing the total number of communications from 6 to 4 per attention layer.


\noindent\ul{\textbf{Principle 4: Gradient Averaging Technique to Synchronize GPUs and Avoid Concatenation.}}
There are two issues from Principles 1 and 3. First, since sequence parallel GPU trains on the same model parameters but using different input sequence segments, the gradients for the model parameters are different for each GPU. The second issue is that the self-attention communication frequency needs to be further reduced to achieve even better scalability and parallel efficiency.

To address both issues, we use a gradient averaging technique to synchronize model parameters and avoid the concatenation for the GPUs' individual self-attention outputs. Therefore, communication frequency is reduced from 4 to 2 per attention layer.
Figs.~\ref{fig:LSS-Transformer}(iii)-(v) use a 2 GPU example to demonstrate how this gradient averaging technique is applied. In the forward pass for the self-attention in Fig.~\ref{fig:LSS-Transformer}(iii), a distributed query $Q_i$ is computed from the input sequence segment $x_i$. Meanwhile, the self-attention input segments are gathered among GPUs before computing collected $K$ and $V$ vectors using a single all-gather fused communication, as explained before in Principle 3. Subsequent computations and memory storage are all distributed and independently updated in the sequence dimension, producing individual self-attention output for each GPU. 

The individual self-attention outputs, however, are not concatenated across GPUs in Fig.~\ref{fig:LSS-Transformer}(iii). Instead, the LSS Transformer allows each GPU to use its assigned sequence segment and individual self-attention output to compute a partial cross-entropy loss and gradients in the backward pass in Figs.~\ref{fig:LSS-Transformer}(iv) and (v). Note that the backward pass in Fig.~\ref{fig:LSS-Transformer}(v) uses reduce-scatter as the backward operation for the all-gather in the forward pass. Finally, the averaged gradients are computed and used for synchronized model parameter updates before training on the next data batch. One important technical detail to mention is that the averaged gradients are not computed for the positional embeddings, which are distributed parameters across GPUs and should not be synchronized.

To understand why this gradient averaging technique can avoid self-attention concatenation and synchronize model parameters at the same time, let us assume that the predicted sequence output from transformer is $y$ and its true label is $\tilde{y}$. The cross-entropy loss for the whole sequence, denoted as $L(y,\tilde{y})$, equals the average of individual token's loss: $L(y,\tilde{y})=\frac{1}{l_x}\sum_{i=1}^{l_x} L(y_i,\tilde{y_i})$, where $l_x$ is sequence length. According to the gradient summation rule, the gradient of $L(y,\tilde{y})$ with respect to model parameters, denoted as $\grad L(y,\tilde{y})$, equals the averaged gradient of each token's loss: $\grad L(y,\tilde{y})=\frac{1}{l_x}\sum_{i=1}^{l_x} \grad L(y_i,\tilde{y_i})$. Therefore, there is no need to concatenate individual self-attention outputs to compute the loss and gradients for the whole sequence. Instead, each GPU uses its scattered individual self-attention output to compute a partial loss and gradient for each sequence segment, before averaging each segment's gradients for a synchronized model parameters update.

By avoiding the expensive concatenation operations in each attention layer, LSS Transformer lowers its communication frequency to only twice per attention layer (one all-gather in forward pass and one reduce-scatter in backward pass) since the gradient averaging occurs only once per data batch. This results in much better scalability and reduced communications compared to other sequence parallel methods.

\begin{figure}
\centering
\includegraphics[width=.85\linewidth]{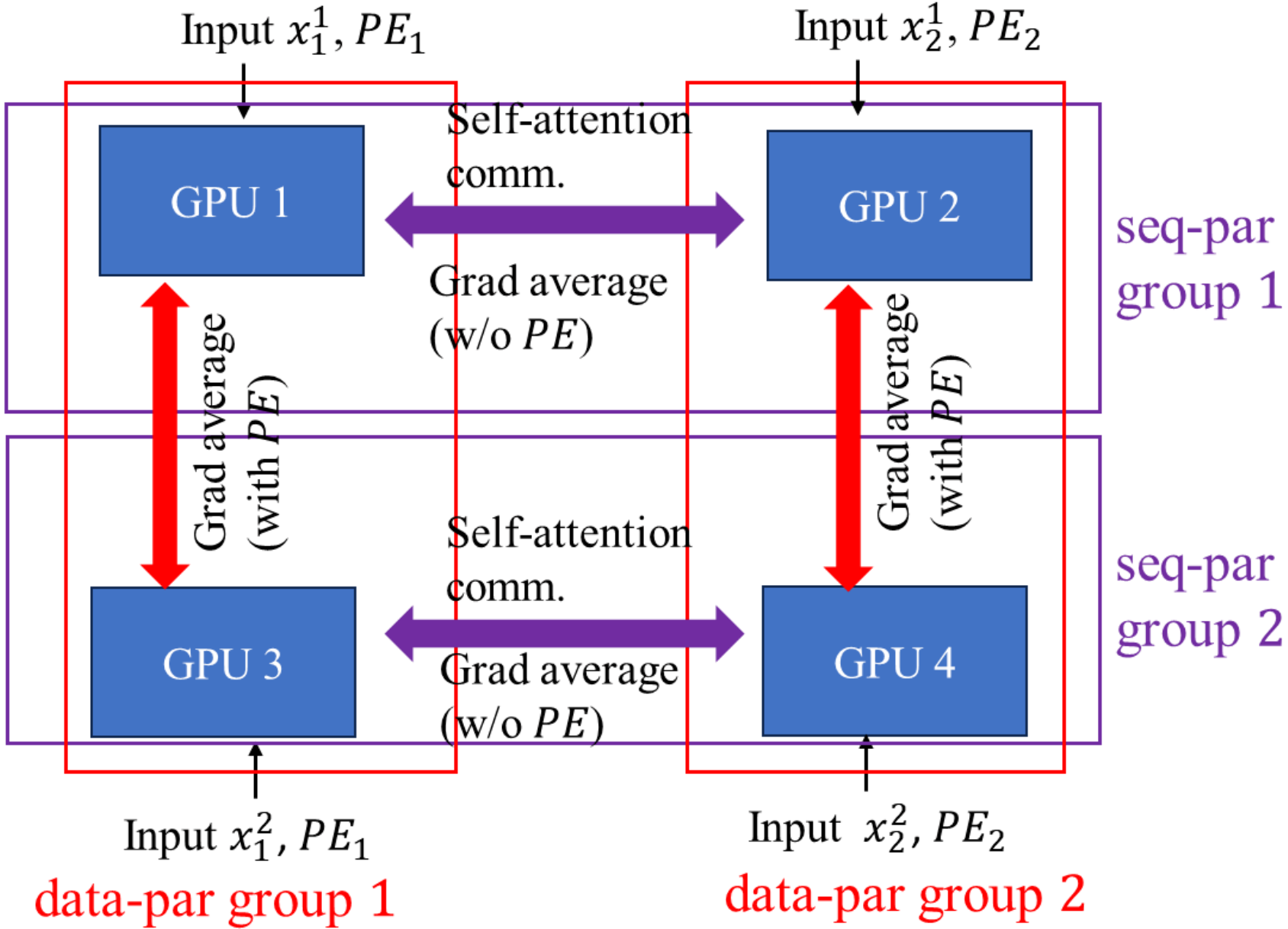}
 \caption{Integrated sequence and data parallelisms with double gradient averaging. The horizontal direction gradient averaging synchronizes parameters without positional embeddings, and the vertical direction gradient averaging includes positional embeddings.}
\label{fig:seq-data-orthogonal} 
\end{figure}

\section{Integrated Sequence \& Data Parallelism}
\label{sec:seq-data-parallel}

The LSS Transformer's sequence parallelism has three limitations. First, it still requires 2 global inter-GPU communications per attention layer, which degrades parallel efficiency at many GPUs. Second, while sequence parallelism tackles the long sequence issue, it does not address computation challenge for training large dataset. Three, sequence parallelism is only one source of parallelism. To scale to a large supercomputer for training, the LSS Transformer needs more sources of parallelism to achieve better scalability.
To address these issues, this section introduces a method to integrate the LSS Transformer's sequence parallelism  with data parallelism. With the integration, the parallel algorithm can (1) achieve better scalability; (2) simultaneously tackle long sequence and large dataset challenges; and (3) constrain the self-attention communications among local communicative groups for reduced overhead.

Despite that sequence and data parallelisms are mostly orthogonal, one technical challenge to overcome is that both parallelisms require model parameter synchronization, but among GPUs in different communicative groups and communicate in different ways. Sequence parallelism requires model parameter synchronization among sequence parallel GPUs, but excludes positional embedding parameters from synchronization given that positional embeddings are distributed in sequence dimension. Data parallelism requires model parameter synchronization among data parallel GPUs, but must include positional embeddings given that data parallel GPUs have the same copy of the positional embedding parameters, but train them 
with different data batches.

To address this issue, we use an innovative double gradient averaging technique to avoid synchronization conflicts for positional embeddings. Fig.~\ref{fig:seq-data-orthogonal}. illustrates an example of how the integrated sequence and data parallelism uses double gradient averaging. In this example, GPUs 1 and 2 process a sequence $x^1$ together using sequence parallelism, with the first segment $x^1_1$ assigned to GPU 1 and the second segment $x^1_2$ assigned to GPU 2. The positional embedding parameters are distributed in the same way with the first half $PE_1$ assigned to GPU 1 and the second half $PE_2$ assigned to GPU 2. Similarly, GPUs 3 and 4 handle a difference sequence $x^2$ using sequence parallelism. 

\begin{table*}
\caption{
The LSS Transformer and Nvidia baseline sequence parallelism's weak scaling experiment for the small model experiment, using only 1 data parallel group.}
\begin{subtable}{.56\linewidth}\centering
\caption{LSS Transformer, data parallel group = 1}
{\small
\begin{tabular}{lcccccr}
\toprule
Nodes & 1   & 6 & 18 & 54 &  144 \\
\midrule
GPUs & 6 & 36 & 108 & 324 &  864 \\
Sequence Groups & 6 & 36 & 108 & 324 &  864 \\
Sequence Length & 348 & 2088 & 6264 & 18792 & 50112 \\
Per GPU Mem. (GB) & 0.54 & 1.01 & 2.05 & 5.94 & 13.58
 \\
FLOPS (x$10^{12}$ flop/s) & 8 & 189 & 881 & 3000 & 8245
 \\
Self-Attn Comp Incr. & 
1 & 6 & 18 & 54 &  144
  \\
Parallel Efficiency & 100\% & 165\% & 174\% & 173\% & 151\%
 \\
\bottomrule 
\end{tabular}
}
\end{subtable}
\begin{subtable}{.43\linewidth}\centering
\caption{Baseline, data parallel group = 1}
{\small
\begin{tabular}{lcccr}
\toprule
Nodes & 1  & 6 & 18  \\
\midrule
GPUs & 6 & 36 & 108  \\
Sequence Groups & 6 & 36 & 108 \\
Sequence Length & 348 & 2088 & 6264  \\
Per GPU Mem. (GB) & 0.94 & 10.29 & OOM \\
FLOPS (x$10^{12}$ flop/s) & 5 & 32 & OOM
 \\
Self-Attn Comp Incr. & 
1 & 6 & OOM
   \\
Parallel Efficiency & 
100\% & 42\% & OOM
 \\
\bottomrule
\end{tabular}
}
\end{subtable}
\label{table:performance-small-model}
\end{table*}

\begin{table*}
\caption{The LSS Transformer and Nvidia baseline weak scaling for the small model experiment, using 4 data parallel groups.}
\begin{subtable}{.56\linewidth} \centering
\caption{LSS Transformer, data parallel groups = 4}
{\small
\begin{tabular}{lcccccr}
\toprule
Nodes & 4 & 24 & 72 & 216 &  576
 \\
\midrule
GPUs & 24 & 144 & 432 & 1296 &  3456
 \\
Sequence Groups & 6  & 36 & 108 & 324 & 864 \\
Sequence Length & 348  & 2088 & 6264 & 18792 &  50112 \\
Per GPU Mem. (GB) & 0.54 & 1.01 & 2.11 & 5.94  & 13.58
 \\
FLOPS (x$10^{12}$ flop/s) & 28 & 703 & 3319 & 10987 & 32784
 \\
Self-Attn Comp Incr. & 
1 & 6 & 18 & 54  & 144
  \\
Parallel Efficiency & 100\% & 167\% & 173\% & 159\% &  161\%
 \\
\bottomrule
\end{tabular}
}
\end{subtable}
\begin{subtable}{.43\linewidth}\centering
\caption{Baseline, data parallel groups = 4}
\centering
{\small
\begin{tabular}{lccr}
\toprule
Nodes & 4 & 24 & 72
 \\
\midrule
GPUs & 24 & 144 & 432
 \\
Sequence Groups & 6 & 36 & 108 \\
Sequence Length & 348 & 2088 & 6264\\
Per GPU Memory (GB) & 0.94 & 10.29 & OOM
 \\
FLOPS (x$10^{12}$ flop/s) & 18 & 126 & OOM
 \\
Self-Attn Comp Incr. & 
1 & 6 & OOM
  \\
Parallel Efficiency & 100\% & 42\% & OOM
 \\
\bottomrule
\end{tabular}
}
\end{subtable}
\label{table:performance-small-model2}
\end{table*}

All GPUs that process the same sequence form a sequence parallel group and each group is shown as a horizontal purple box in Fig.~\ref{fig:seq-data-orthogonal}. Each sequence parallel group's cross-GPU communications, shown as a horizontal purple arrow, are local and confined among GPUs in the same sequence parallel group. These communications involve a fused all-gather and a reduce-scatter in each attention layer for computing distributed self-attention. In addition, a gradient averaging is required once per data batch for model parameters synchronization and avoiding self-attention output concatenation, as discussed before in Sec.~\ref{sec:LSS Transformer}. The positional embedding parameters, however, are excluded from gradient averaging, as they are distributed across sequence parallel GPUs, and should not be synchronized. 

Meanwhile, GPUs 1 and 3 are in a data parallel group, shown as a vertical red box in the figure, and GPUs 2 and 4 are in another data parallel group. GPUs in the same data parallel group process sequence segments from different data batches, but these sequence segments share the same spatial position within their sequences, thereby sharing the same positional embedding. The only needed inter-GPU communications in the same data parallel group, shown as vertical red arrows in the figure, are gradient averaging to synchronize parameters trained with different batches. Similar to sequence parallel groups, the communication for the data parallel groups is also localized and confined within each group. The gradient averaging for data parallelism, however, must include positional embeddings for synchronization, given that the training segments in the same data parallel group share the same positional embedding.

\section{Results}
\label{sec:experiments}
\subsection{Experiment Setup}
\label{subsec:dataset_platform}

\textit{\textbf{Dataset:}} \textit{enwik8} is a popular 100-million-byte Wikipedia XML character training dataset~\cite{hutter-prize,Beltagy2020Longformer}. Initially used as the testing dataset for the Hutter Prize, the dataset can be downloaded at~\cite{enwik8-download} for public benchmark evaluation with a score board available at~\cite{enwik8-paper-list}.

\noindent\textit{\textbf{Computing platform:}} 
Experiments were conducted on Oak Ridge National Lab's Summit supercomputer, which has 6 NVIDIA V100 GPUs and 2 POWER9 CPUs for each node.
Nodes are connected via Mellanox EDR 100G InfiniBand Non-blocking Fat Tree network.
Each POWER9 CPU in the node is densely-connected to 3 GPUs with Nvidia NVlinks, where each link has 100 GB/s bidirectional bandwidth, and the two CPUs for each node are connected via an X bus with 64 GB/s bidirectional
bandwidth. Each CPU has 22 cores (4 hardware threads for each) and 256 GB DRAM memory. Each GPU has 80 streaming multiprocessors and 16 GB memory. There are additional 54 ``high memory” nodes, which has 32 GB of memory per GPU.

\noindent\textit{\textbf{Software}}: The LSS Transformer is developed in PyTorch and will be made publicly available in the next revision.

\subsection{Small Model Experiment}
\textit{\textbf{Model:}} All experiments in the small model experiment subsection trains a decoder-only transformer (GPT) with a 512 embedding size, 6 attention layers and 8 multi-heads for a total of 20 million parameters. The input data batch size is 4. We choose this model size for two reasons. First, this is the standard model for the \textit{enwik8} benchmark evaluation with an excellent bits-per-character accuracy score at 1.0~\cite{Beltagy2020Longformer,Al-Rfou_2019,sukhbaatar-2019}. Second, choosing a small model size allows us to maximize memory usage and evaluate performance for scaling long sequences, instead of maximizing memory for storing parameters for a large model.

\noindent\textit{\textbf{Sequence Parallelism Weak Scaling:}} 
Tables~\ref{table:performance-small-model}(a) and~\ref{table:performance-small-model}(b) are the weak scaling performance comparison between the LSS Transformer and Nvidia baseline sequence parallelism. We increased both the sequence lengths and the number of sequence parallel GPUs at the same rate, while keeping the number of data parallel group to be 1.

\begin{figure*}
\centering
\begin{subfigure}{.33\linewidth}
\includegraphics[width=\linewidth]{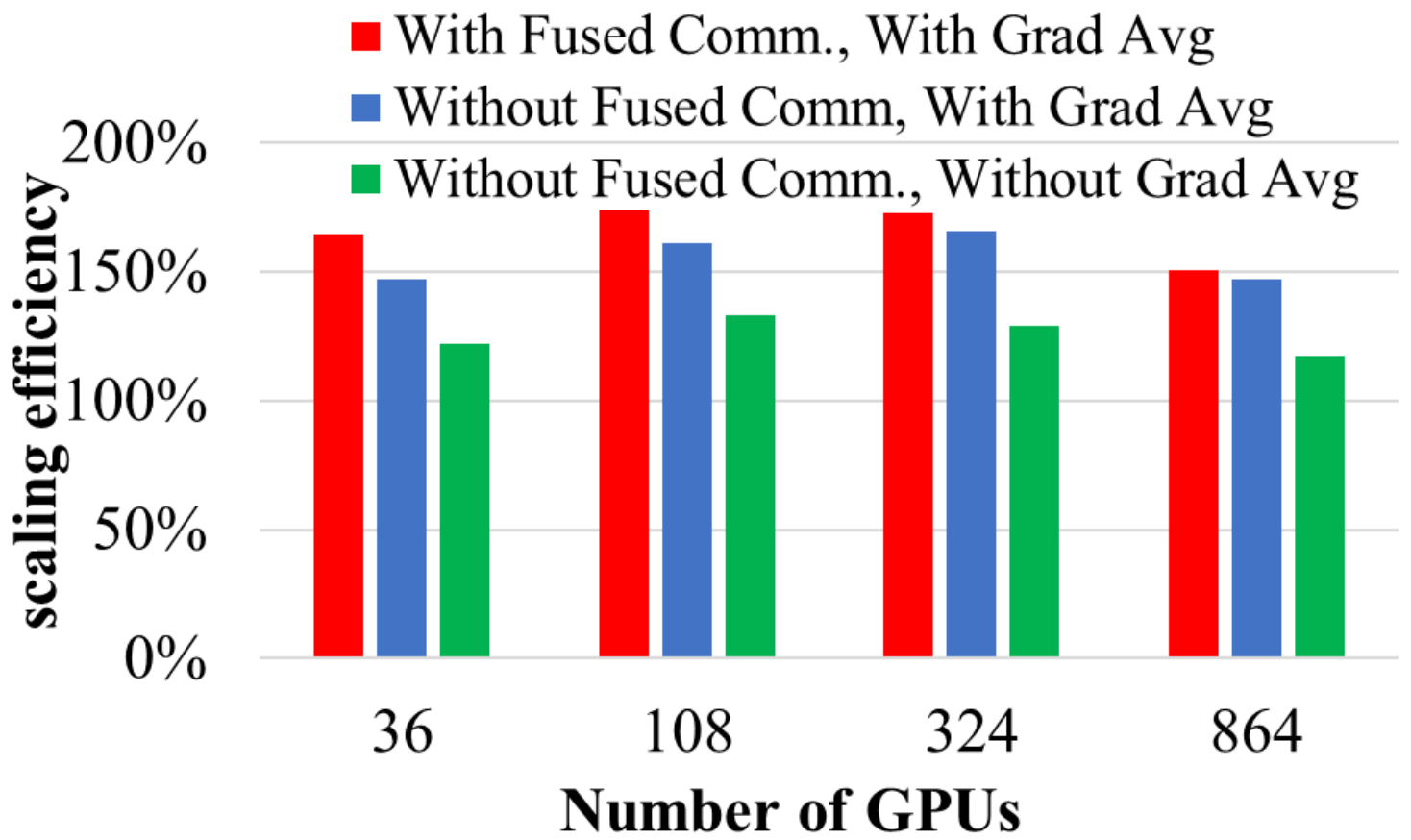}
 \caption{Scaling efficiency with and without fused communication and gradient averaging}
\end{subfigure}
\begin{subfigure}{.33\linewidth}
\includegraphics[width=\linewidth]{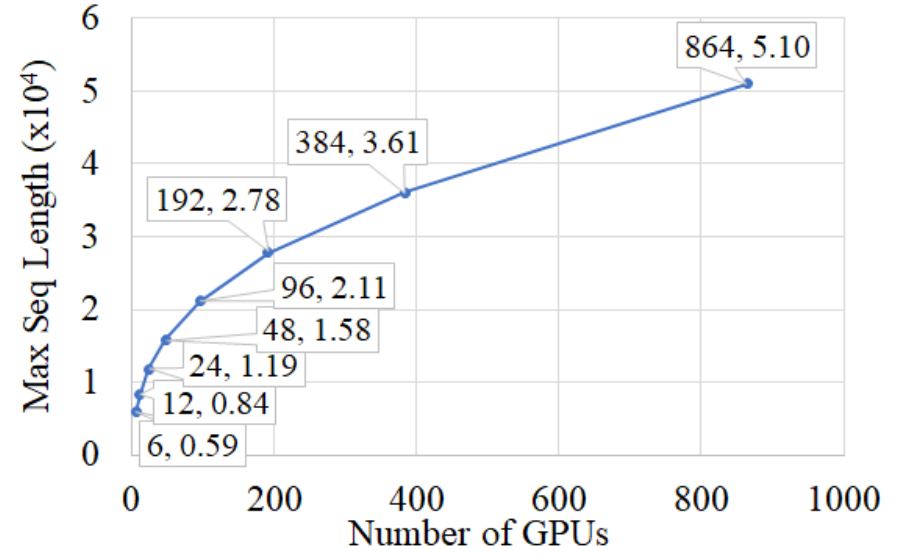}
 \caption{Maximum sequence length}
\end{subfigure}
\begin{subfigure}{.33\linewidth}
\includegraphics[width=\linewidth]{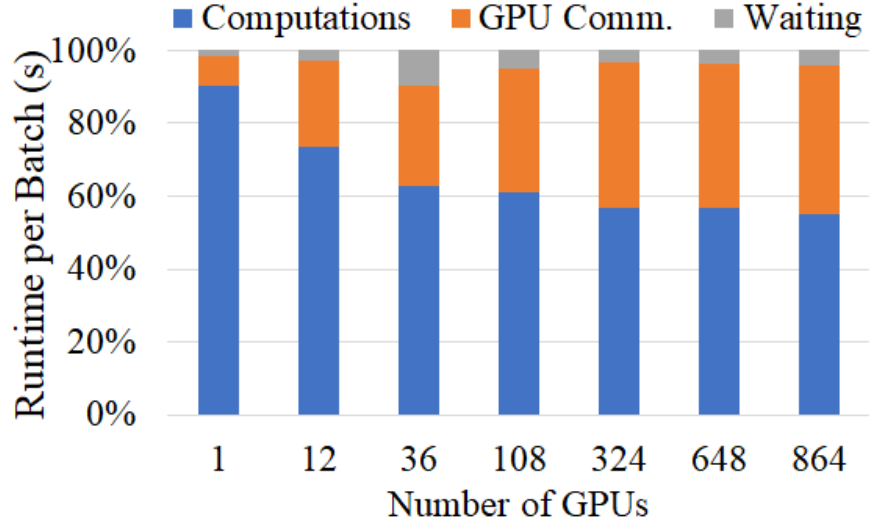}
 \caption{Runtimes breakdown for Table~\ref{table:performance-small-model}(a)}
\end{subfigure}
\caption{
(a) Scaling efficiency with and without fused communication and gradient averaging. (b) maximum sequence length at different numbers of sequence parallel GPUs. (c) the runtime breakdown for the LSS Transformer's small model experiment weak scaling with only 1 data parallel group.}
\label{fig:experiment-analysis}
\end{figure*}

\begin{table*}
\caption{
Performance comparison between the two algorithms on the small model, using only 1 data parallel group. Sequence lengths increase sub-linearly with a growth rate proportional to the square root of the number of GPUs.
}
\begin{subtable}{.56\linewidth}\centering
\caption{LSS Transformer, data parallel group = 1}
{\small
\begin{tabular}{lcccccr}
\toprule
Nodes & 1   & 6 & 18 & 54 &  144 \\
\midrule
GPUs & 6 & 36 & 108 & 324 &  864 \\
Sequence Groups & 6 & 36 & 108 & 324 &  864 \\
Sequence Length & 366 & 900 & 1512 & 2592 & 4320 \\
Per GPU Mem. (GB) & 0.53 & 0.62 & 0.74 & 0.89 & 1.15
 \\
FLOPS (x$10^{12}$ flop/s) & 8 & 83 & 266 & 673 & 1280
 \\
Self-Attn Comp Incr. & 
1 & 2 & 4 & 6 &  10
  \\
Parallel Efficiency & 100\% & 120\% & 114\% & 96\% & 72\%
 \\
\bottomrule 
\end{tabular}
}
\end{subtable}
\begin{subtable}{.43\linewidth}\centering
\caption{Baseline, data parallel group = 1}
{\small
\begin{tabular}{lcccr}
\toprule
Nodes & 1  & 6 & 18  \\
\midrule
GPUs & 6 & 36 & 108  \\
Sequence Groups & 6 & 36 & 108 \\
Sequence Length & 366 & 900 & 1512  \\
Per GPU Mem. (GB) & 0.89 & 2.55 & 5.86 \\
FLOPS (x$10^{12}$ flop/s) & 5 & 28 & 71
 \\
Self-Attn Comp Incr. & 
1 & 2 & 4
   \\
Parallel Efficiency & 
100\% & 56\% & 36\%
 \\
\bottomrule
\end{tabular}
}
\end{subtable}
\label{table:performance-sublinear}
\end{table*}

The first two rows of the tables indicate the number of nodes and GPUs, with 6 GPUs per node. The third row represents the number of sequence parallel groups, which is equal to the number of GPUs in this case as the number of data parallel group is 1. The fourth row shows the sequence lengths, increasing proportionally with the number of GPUs. 

The fifth row displays the average per-GPU peak memory footprint in Gigabytes (GBs). At 6 nodes, the LSS Transformer's per-GPU memory footprint is 1.01 GB, and it scales to 144 nodes with 13.58 GB per GPU. Since Transformer has a quadratic memory complexity of $O(l_x^2/N)$, where $l_x$ is sequence length and $N$ is the number of GPUs, increasing sequence length $l_x$ and number of GPUs $N$ at the same rate will still lead to a linear increase of memory. This explains why LSS Transformer has a small memory footprint at 1 node but increases to much larger memory footprint at more nodes.
In comparison, the baseline sequence parallelism has a per-GPU memory footprint of 10.29 GB at 6 nodes, over 10 times larger than that of the LSS Transformer at the same nodes. The baseline sequence parallelism cannot scale beyond 6 nodes due to memory constraint, resulting in "OOM" (out of memory).

The sixth row represents the number of single-precision floating point operations (FLOP) across all GPUs in a second. The LSS Transformer achieves a computation throughput of 8 petaflops at 144 nodes with sequence length of 50,112. In comparison, the baseline sequence parallelism is 5.9 times slower at 6 nodes, achieving a throughput of 32 teraflops, and cannot scale further due to memory constraint.

The seventh row shows the recorded per-GPU computations increase for the self-attention unit relative to the per-GPU computations at 1 node. Since transformer has a cubic computation complexity, distributing computations across GPUs will still lead to computation increase for weak scaling.

The eighth row represents parallel efficiencies, which is the ratio between the actual speedup and the ideal speedup. The LSS Transformer maintains a 151\% super-linear parallel efficiency at 144 nodes, while the baseline sequence parallelism's efficiency drops to 42\% at only 6 nodes.

\noindent\textbf{\textit{Integrated Sequence \& Data Parallelism Weak Scaling:}} To understand how the integration of data and sequence parallelisms accelerates training speed and reduces communication overhead, Table~\ref{table:performance-small-model2}(a) repeats the same experiment as Table~\ref{table:performance-small-model}(a), but with 4 data parallel groups. This means that the total number of GPUs quadruples accordingly, but the numbers of sequence parallel groups remain the same as before. By comparing the results between the 4 data parallel groups and the single data parallel group in Tables~\ref{table:performance-small-model} and~\ref{table:performance-small-model2}, we observe that the FLOPs throughput increases by almost 4 times from 1 to 4 data parallel groups with 4 times more GPUs, achieving 32 petaflops at 3456 GPUs. This result indicates that the proposed local communication scheme enables the integrated sequence and data parallelism with little additional communication overhead and the integration of these two parallelisms is an effective approach for achieving more scalability and faster training.

\begin{table*}
\caption{
Performance comparison between the two algorithms on a large 1.5 billion model, using only 1 data parallel group, and without any model parallelism. Sequence lengths increase with a growth rate proportional to the square root of the number of GPUs.
}
\begin{subtable}{.56\linewidth}\centering
\caption{LSS Transformer, data parallel group = 1}
{\small
\begin{tabular}{lcccccr}
\toprule
Nodes & 1   & 6 & 18   \\
\midrule
GPUs & 6 & 36 & 108   \\
Sequence Groups & 6 & 36 & 108   \\
Sequence Length & 366 & 900 & 1512   \\
Per GPU Mem. (GB) & 21.67  & 22.48 & 23.34  
 \\
FLOPS (x$10^{12}$ flop/s) & 52 & 518 & 2010  
 \\
Self-Attn Comp Incr. & 
1 & 2 & 4  
  \\
Parallel Efficiency & 100\% & 94\% & 92\% 
 \\
\bottomrule 
\end{tabular}
}
\end{subtable}
\begin{subtable}{.43\linewidth}\centering
\caption{Baseline, data parallel group = 1}
{\small
\begin{tabular}{lcccr}
\toprule
Nodes & 1  & 6  \\
\midrule
GPUs & 6 & 36   \\
Sequence Groups & 6 & 36  \\
Sequence Length & 366 & 900   \\
Per GPU Mem. (GB) & 25.28  & OOM    \\
FLOPS (x$10^{12}$ flop/s) & 23 & OOM
 \\
Self-Attn Comp Incr. & 
1 & OOM 
   \\
Parallel Efficiency & 
100\% & OOM 
 \\
\bottomrule
\end{tabular}
}
\end{subtable}
\label{table:performance-large-model}
\end{table*}

\noindent\textbf{\textit{Super-Linear Speedup:}}  The LSS Transformer achieves super-linear scaling in Tables~\ref{table:performance-small-model} and~\ref{table:performance-small-model2} due to two reasons. First, longer sequences result in increased work for each GPU due to the self-attention computation increase, leading to higher GPU utilization rates for longer sequences and super-linear scaling. Measured by PyTorch Cuda utilization report, GPU utilization rate for the LSS Transformer increases from 33\% at 1 node with a sequence length of 348 to 83\% at 6 nodes with a sequence length of 2,088. Second, the LSS Transformer's low communication overhead significantly contributes to its excellent parallel efficiency. Fig.~\ref{fig:experiment-analysis}(a) shows the scaling efficiency with and without the fused communication and gradient averaging techniques, which were both introduced in Sec.~\ref{sec:LSS Transformer}. At 864 GPUs and sequence length of 50,112, the scaling efficiency with both techniques are 151\%. Efficiency with gradient averaging but without fused communication is 147\%, whereas efficiency without either technique is dropped to 118\%.

\noindent\textbf{\textit{Maximal Sequence Length:}} 
Fig.~\ref{fig:experiment-analysis}(b) depicts the maximal sequence length when scaling the number of GPUs while maximizing memory capacity. Each numbered pair in the figure corresponds to the number of GPUs and the maximal sequence length. For example, $(6, 0.59)$ indicates that 6 GPUs can scale up to a maximal sequence length of $0.59 \times 10^4$. We can observe that the maximal sequence length follows a square root curve in the graph. Since transformer has quadratic memory complexity with longer sequences, the maximal sequence length increases asymptotically with square root functions as the total memory capacity grows.

\noindent\textbf{\textit{Runtime Breakdown.}} Fig.~\ref{fig:experiment-analysis}(c) illustrates the runtime breakdown for the weak scaling results presented in Table~\ref{table:performance-small-model}(a), focusing on a single data parallel group. The blue bars represent the percentage of runtime spent on computations, while the orange bars indicate the percentage of runtime for communications. The grey bars denote the GPU waiting times. At 36 GPUs, the communication overhead accounts for 27\% of the total runtime. As the number of GPUs scales to 864 (144 nodes), the communication overhead becomes 40\% of the runtime. 

\noindent\textbf{\textit{Scaling Sub-Linearly Increased Sequence Lengths.}} One way to limit memory and computations increase is to increase the sequence lengths at a sub-linear rate. Table~\ref{table:performance-sublinear} repeats the same scaling experiment
as Table~\ref{table:performance-small-model}, but with the sequence lengths increasing proportionally
to the square root of the number of GPUs. As a result,
both memory and self-attention computations from rows 5 and 7 of the table
also increase at a rate asymptotic to square root
functions. The memory footprint for the
LSS-Transformer is only at 1.15 GB per GPU when scaling to 864
GPUs, and the self-attention computations is only 10 times more
than that for a single node. The LSS-Transformer’s scaling remains highly efficient at 94\% with 864 GPUs. In contrast, scaling efficiency for Nvidia’s
sequence parallelism drops to 31\% at 108 GPUs and cannot
scale further due to memory constraints.

\subsection{Large Model Experiment}
Table~\ref{table:performance-large-model} repeats the same experiment as Table~\ref{table:performance-sublinear}, but trains a large 1.5-billion-parameter GPT model that has a 2048 embedding size, 24 attention layers and 16 multi-heads. experiments are run on the high-memory nodes of Summit with 32 GBs of memory per GPU, and no model parallelism is used for this experiment. Since most of the memory capacity is now used for storing model parameters instead of long sequences, we can notice that all runs in this table use much more memory than the small model experiment.
LSS Transformer maintains a high scaling efficiency of 92\% at 108 GPUs and a smaller memory footprint compared to the baseline parallelism. In contrast, the baseline parallelism cannot scale beyond a single node due to memory constraint an its FLOPs throughput is 2.3 times smaller than that for LSS Transformer at a single node.

\section{Conclusion}

This paper introduced the Long-Short Sequence Transformer (LSS Transformer), a novel algorithm and a general framework for distributing long sequence in transformer models. It uses a novel distributed self-attention mechanism, along with fused communication and double gradient averaging techniques, to achieve impressive speedups and memory reduction with minimal communication overhead.

In conclusion, the LSS Transformer is a significant step forward for addressing transformer's long sequence problem. We believe that our approach provides an important contribution to the research field and enables ultra-long sequence training, especially to applications that benefit from long-range token dependencies, such as DNA sequence analysis, long document summary, and imaging applications.

\bibliographystyle{mlsys2024}
\bibliography{main}

\end{document}